\documentclass[aps,twocolumn,prl,showpacs,preprintnumbers,amsmath,amssymb]{revtex4}
\usepackage{graphicx}% Include figure files
\usepackage{dcolumn}% Align table columns on decimal point
\usepackage{bm}% bold math
\usepackage{txfonts}

\begin{document}
\preprint{APS/123-QED}

\title{Divergence of the Magnetic Gr\"{u}neisen Ratio at the Field-Induced Quantum Critical Point in YbRh$_2$Si$_2$}% Force line breaks with \\

\author{Y. Tokiwa$^{1,2}$}\thanks{present address: I. Physik.
Institut, Georg-August-Universit\"{a}t G\"{o}ttingen, D-37077
G\"{o}ttingen}
\author{T. Radu$^{1}$}\thanks{present address: Faculty of
Physics, Babes Bolyai University, Cluj-Napoca, Romania}

\author{P. Gegenwart$^{3}$}
\author{C. Geibel$^{1}$}
\author{F. Steglich$^{1}$}
\affiliation{$^{1}$Max-Planck Institute for Chemical Physics of
Solids, D-01187 Dresden, Germany\\$^{2}$Los Alamos National
Laboratory, Los Alamos, New Mexico 87545, USA\\$^{3}$I. Physik.
Institut, Georg-August-Universit\"{a}t G\"{o}ttingen, D-37077
G\"{o}ttingen}

\date{\today}% It is always \today, today,

\begin{abstract}
The heavy-fermion metal YbRh$_2$Si$_2$ is studied by low-temperature
magnetization $M(T)$ and specific-heat $C(T)$ measurements at
magnetic fields close to the quantum critical point ($H_c=0.06$~T,
$H\perp c$). Upon approaching the instability, $dM/dT$ is more
singular than $C(T)$, leading to a divergence of the magnetic
Gr\"uneisen ratio $\Gamma_{\rm mag}=-(dM/dT)/C$. Within the Fermi
liquid regime, $\Gamma_{\rm mag}=-G_r(H-H_c^{fit})$ with
$G_r=-0.30\pm 0.01$ and $H_c^{fit}=(0.065\pm 0.005)$~T which is
consistent with scaling behavior of the specific-heat coefficient in
YbRh$_2$(Si$_{0.95}$Ge$_{0.05}$)$_2$. The field-dependence of
$dM/dT$ indicates an inflection point of the entropy as a function
of magnetic field upon passing the line $T^\star(H)$ previously
observed in Hall- and thermodynamic measurements.
%The magnetic Gr\"{u}neisen ratio $\Gamma_{\rm mag}=-(dM/dT)/C$ for
%the heavy fermion compound YbRh$_2$Si$_2$ is determined by specific
%heat $C(T)$ and magnetization $M(T)$ measurements. At the quantum
%critical field $H$=0.06\,T, a more singular temperature dependence
%in $-(dM/dT)/T$ than that in $C/T$ results in a diverging
%$\Gamma_{\rm mag}$ as $T\rightarrow0$. The saturated $\Gamma_{\rm
%mag}$($T\rightarrow0$) for the Fermi liquid region at fields above
%$H_c$ follows the theoretical prediction $(H-H_c^{fit})^{-1}$,
%yielding $H_c^{fit}$=0.065\,T, in a good agreement with the quantum
%critical field 0.06\,T ($H\perp c$). We also found an anomalous
%region below $T\sim$0.25\,K and at fields 0.1\,T$<H<$0.2\,T, where
%$-(dM/dT)/C$ still diverges while $C/T(T)$ is already saturated.

\end{abstract}
\pacs{71.10.HF,71.27.+a}
\maketitle

Quantum criticality is a topic of extensive current research
interest in condensed matter physics because it results in unusual
finite-temperature properties and promotes the formation of novel
states like unconventional superconductivity (for recent reviews
see~\cite{Stewart,Lohneysen_Review,Nature physics}). A quantum phase
transition occurs at $T=0$ upon tuning an external parameter $r$
like pressure, doping or magnetic field to a critical value $r_c$.
For a continuous (second-order) phase transition at $r_c$, a quantum
critical point (QCP) emerges, which has wide reaching influence on
the temperature dependence of thermodynamic and transport
properties. In metallic systems near magnetic QCPs pronounced
non-Fermi liquid (NFL) effects have been observed, which depend on
the dimensionality (D), type of magnetic interaction, i.e.
antiferromagnetic (AF) or ferromagnetic (FM), as well as type of
quantum criticality (itinerant, locally critical, Kondo-breakdown,
etc.)~\cite{Millis,Moriya,Si,Coleman,Senthil04,Pepin}. Such NFL
effects are related to an anomalous enhancement of the entropy
$S(T,r)$ near the QCP, resulting from quantum critical fluctuations.
For a {\it pressure-tuned} QCP ($r\propto p-p_c$, $p_c$: critical
pressure) it has been pointed out, that the Gr\"uneisen ratio
$\Gamma \propto \alpha/C$ of thermal expansion $\alpha$ to specific
heat $C$ diverges in the approach of the
QCP~\cite{ZhuLijun:UnidGp,GarstM:SigctG}, as recently found for
various
systems~\cite{KuchlerR:DivtGr,KuchlerR:Grurdq,Kuchler06,Lorenz07}.
When the control parameter to tune the system to the QCP is the
magnetic field ($r=H-H_c$, $H_c$: critical magnetic field), a
corresponding divergence has been predicted for the {\it magnetic}
Gr\"uneisen parameter $\Gamma_{mag}=-(dM/dT)/C$
\cite{GarstM:SigctG}. This property equals the magnetocaloric
effect, i.e is proportional to the slope of isentropes in the
temperature vs. magnetic field phase diagram \cite{GarstM:SigctG}.
Below, we demonstrate the use of magnetization for characterizing a
NFL and report for the first time a divergence of $\Gamma_{mag}$ in
the approach of a (field-induced) QCP.

We focus on tetragonal YbRh$_2$Si$_2$, which is a clean and
stoichiometric heavy-fermion metal that displays a magnetic
field-tuned QCP at $H_c=0.06$~T (0.66~T) for $H\perp c$ ($H\parallel
c$) \cite{GegenwartP:Magiqc}. This QCP arises when very weak AF
ordering at $T_N=70$~mK, with an ordered moment as small as $2\times
10^{-3}\mu_B$/Yb \cite{Ishida}, is continuously suppressed by
magnetic field. Various thermodynamic, magnetic and transport
experiments on YbRh$_2$Si$_2$ as well as its slightly Ge-doped
variant YbRh$_2$(Si$_{1-x}$Ge$_x$)$_2$ ($T_N=20$~mK, $H_c=0.027$~T,
$H\perp c$) have revealed evidence for quantum criticality, that is
controlled by magnetic
field~\cite{GegenwartP:Magiqc,CustersJ:brehea}: At $H=H_c$, $C(T)/T$
exhibits a stronger than logarithmic divergence while $\rho$ has
$T$-linear temperature dependence. For $H>H_c$, Fermi-liquid (FL)
behavior is induced, as evidenced from the electrical resistivity,
described by $\rho(T)=\rho_0+AT^2$, with the coefficient $A(H)$
diverging towards $H_c$ proving a field-induced
QCP~\cite{GegenwartP:Magiqc,CustersJ:brehea}. The Sommerfeld
coefficient $\gamma(H)$ in the FL regime ($H>H_c$) displays a
$(H-H_c)^{-1/3}$ divergence, which is incompatible with the
itinerant theory for a spin-density-wave QCP~\cite{CustersJ:brehea}.
Bulk susceptibility~\cite{GegenwartP:FM,TokiwaY:Fiesth} and nuclear
magnetic resonance experiments~\cite{IshidaK:YbRsft} indicate that
in a wide regime of the $T-H$ phase diagram the critical
fluctuations have a FM character. Only for temperatures below 0.4~K
and fields below 0.25~T, i.e. very close to the AF ordered phase, AF
fluctuations dominate. For $H>H_c$, a strongly enhanced
Sommerfeld-Wilson ratio that exceeds a value of 30 upon approaching
the critical field has been observed~\cite{GegenwartP:FM}. Note,
however, that the observed temperature dependence of the bulk
susceptibility and spin-lattice relaxation rate in the quantum
critical regime cannot be explained within the itinerant theory for
FM quantum critical fluctuations in either 2D or
3D~\cite{GegenwartP:Mulesa}.

The critical Gr\"uneisen ratio $\Gamma_{cr}=V_{mol}/\kappa_T\times
\alpha_{cr}/C_{cr}$ ($V_{mol}$: molar volume, $\kappa_T$: isothermal
compressibility), where $\alpha_{cr}$ and $C_{cr}$ denote the volume
thermal expansion and specific heat after subtraction of
non-critical contributions~\cite{ZhuLijun:UnidGp} has been studied
for YbRh$_2$(Si$_{1-x}$Ge$_x$)$_2$ at zero magnetic field and
temperatures down to 80~mK~\cite{KuchlerR:DivtGr}. Below 0.6~K, a
$\Gamma_{cr}\propto T^{-0.7}$ divergence has been obtained, which
may indicate a local type of QCP~\cite{KuchlerR:DivtGr}. Indeed,
detailed studies of the evolution of the Hall coefficient upon
field-tuning the system suggest a drastic change of the Fermi volume
due to a localization of the $4f$-electrons at the
QCP~\cite{PaschenS:Haleah}. Furthermore, thermodynamic evidence for
an additional energy scale $T^\star(H)$, possibly related to the
Kondo-breakdown was obtained from magnetostriction and magnetization
experiments~\cite{GegenwartP:Mulesa}. Below, we address the nature
of quantum criticality and the evolution of the entropy across
$T^\star(H)$ by means of the {\it magnetic} Gr\"uneisen ratio, which
is the appropriate thermodynamic property for a field-induced QCP.

For our measurements, we used high-quality single crystals ($\rho_0=1~\mu\Omega$cm) characterized before~\cite{GegenwartP:Magiqc}. The magnetization
was measured utilizing a high-resolution capacitive Faraday magnetometer~\cite{SAKAKIBARAT:Farfmh}. Specific-heat measurements have been performed
with the quasi-adiabatic heat-pulse technique.

\begin{figure}[t]
\includegraphics[width=0.8\linewidth,keepaspectratio]{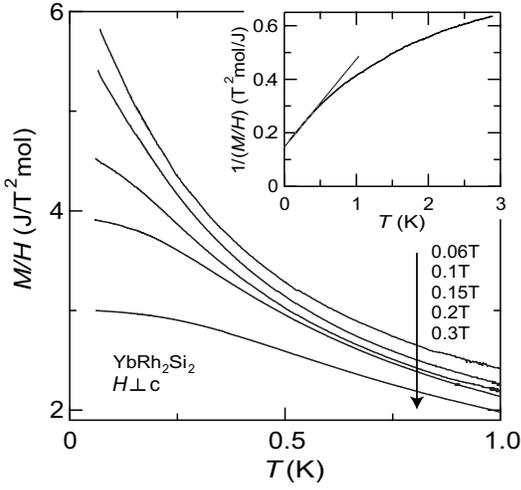}
\caption{Magnetization divided by field $M/H$ of YbRh$_2$Si$_2$ as a
function of temperature. Inset: Inverse of $M/H$ vs temperature.
Dashed line indicates linear fit for low-temperature region
($T<0.3$~K).}
\end{figure}

Figure~1 shows the magnetization divided by field $M/H$ as a
function of temperature. For fields above 0.1\,T, $M(T)/H$ tends to
saturate, while it keeps increasing down to the lowest temperature
below 0.1\,T. The strong increase of $M(T)/H$ with decreasing
temperature at the critical field $H_c$=0.06\,T is reflected in the
negative curvature in the inverse of $M/H$ (see the inset). The
deviation from the Curie-Weiss law above 0.3\,K may be due to FM
correlations in this system. Linear fitting of the data in the
low-temperature region ($T<0.3$\,K) yields a Weiss temperature
$\theta=-0.46$\,K and an effective magnetic moment $\mu_{\rm
eff}$=1.6\,$\mu_B$/Yb$^{3+}$ ion, close to the reported values at
zero field in a previous ac-susceptibility
study~\cite{GegenwartP:Magiqc}.
%The absolute value of $\theta$ decreases upon $H$ decreasing towards
%$H_c$ (not shown).
At $H>H_c$, $M(T)/H$ saturates and does not reach zero. At a FM QCP,
the magnetization is expected to diverge as $T\rightarrow$0 with
$\theta$=0. The finite negative $\theta$ at $H_c$ indicates that AF
fluctuations dominate close to the QCP~\cite{IshidaK:YbRsft}.

\begin{figure}[t!h]
\includegraphics[width=1.0\linewidth,keepaspectratio]{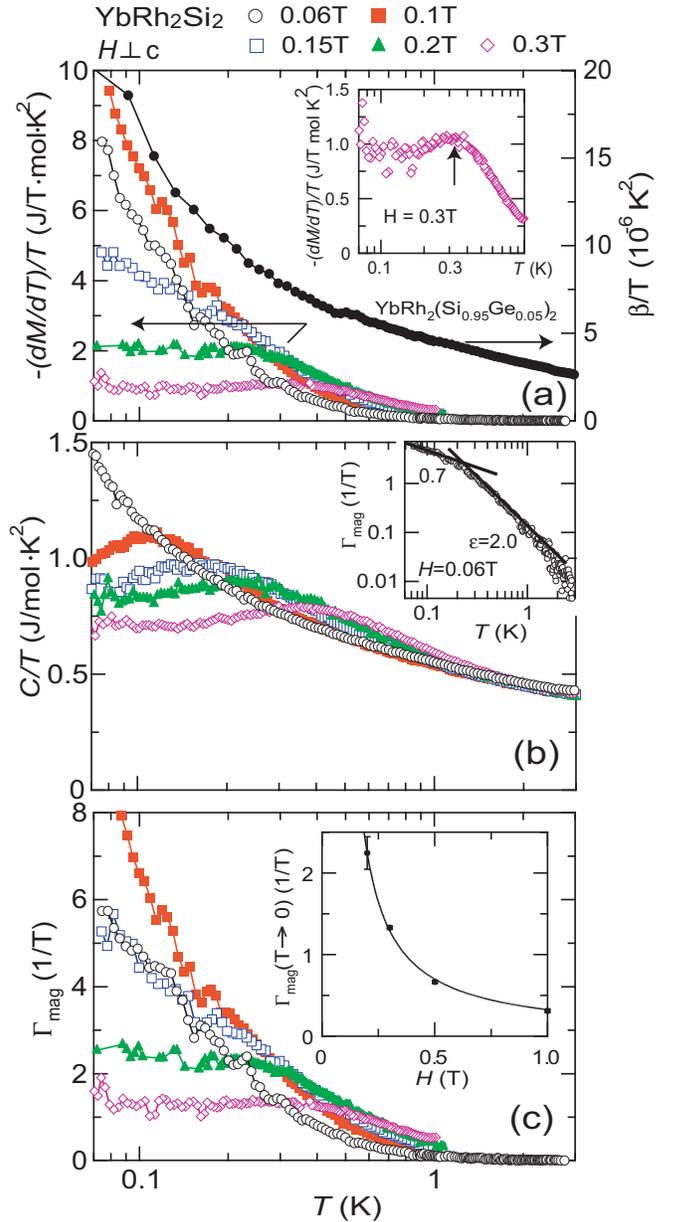}
\caption{(Color online) (a) Temperature derivative of the
magnetization as $-(dM/dT)/T$ (left axis), (b) electronic specific
heat as $C(T)/T$ and (c) magnetic Gr\"{u}neisen ratio $\Gamma_{\rm
mag}$ vs temperature (on a logarithmic scale) for YbRh$_2$Si$_2$ in
magnetic fields applied perpendicular to $c$-axis. Volume thermal
expansion of YbRh$_2$(Si$_{0.95}$Ge$_{0.05}$)$_2$ as $-\beta/T$ at
zero field is shown for comparison in (a) (right
axis)~\cite{KuchlerR:DivtGr}. Inset in (a): Expanded plot for
$-(dM/dT)/T$ vs $T$ for $H$=0.3\,T. The arrow indicates the position
of the maximum. Inset in (b): $\Gamma_{\rm mag}$ vs temperature for
YbRh$_2$Si$_2$ on a log-log plot. The solid lines represent
$\Gamma_{\rm mag}(T)\propto T^{-\epsilon}$ with $\epsilon$=0.7 and
2.0 for low- and high-temperature regions, respectively. Inset in
(c): Field dependence of saturated $\Gamma_{\rm mag}$, as
$T\rightarrow0$. The solid line indicates $-G_r(H-H_c^{fit})^{-1}$
with $G_r=-0.3$ and $H_c^{fit}$=0.065\,T.}
\end{figure}

The temperature derivative of the magnetization for various
different magnetic fields at and above the critical field has been
investigated between 70~mK and 3~K. In Figure 2, we compare the
temperature dependence of  $-(dM/dT)/T$, displayed in part (a) with
that of the respective specific heat coefficient $C(T)/T$ in part
(b) and analyze the ratio between both quantities, $\Gamma_{mag}$
(Fig. 2c). At the critical field $H_c=0.06$~T, both quantities
diverge upon cooling, but the divergence in $-(dM/dT)/T$ is much
stronger, resulting in a divergent magnetic Gr\"{u}neisen ratio. At
$H=0.1$~T, $-(dM/dT)/T$ diverges even stronger than at 0.06~T in the
investigated temperature regime. However, the analysis of the
field-dependence of the magnetic Gr\"{u}neisen ratio, discussed
below, is consistent with a QCP at 0.06~T and suggests a saturation
of the 0.1~T data due to a crossover from NFL to FL behavior at
temperatures below 70~mK. Such crossover scales are known to be
different for different physical quantities \cite{Moriya}. Indeed,
the specific heat coefficient $C(T)/T$ displays such a NFL to FL
crossover for $H=0.1$~T already around 100~mK, whereas the
respective crossover in electrical resistivity at the same field is
located at 80~mK \cite{GegenwartP:Magiqc}.

In Fig.~2(a) (right axis), we also show the volume thermal expansion
as $\beta/T$ for
YbRh$_2$(Si$_{0.95}$Ge$_{0.05}$)$_2$~\cite{KuchlerR:DivtGr} for
comparison. The divergence of $-(dM/dT)/T$ is stronger at high
temperatures and becomes comparable to $\beta/T$ at low
temperatures. The ratio between $\beta$ and $dM/dT$ gives the
pressure dependence of the critical field as expressed by
$dH_c/dp$=$V_m\beta/(dM/dT)$, where $V_m$ is the molar
volume~\cite{GarstM:SigctG}. This ratio is temperature dependent and
decreases with decreasing temperature (not shown). At low
temperatures it tends to saturate and the extrapolated value of
$V_m\beta/(dM/dT)$ for $T\rightarrow 0$ is $(0.19\pm 0.05)$\,T/GPa
which is consistent with previous hydrostatic pressure
experiments~\cite{TokiwaY:Fiesth}.
%Interestingly, at 0.15~T the Knight shift, being proportional to the
%{\bf q}=0 susceptibility, saturates below 200~mK whereas $1/T_1T$,
%probing the {\bf q}-averaged dynamical susceptibility, still
%diverges down to 70~mK. This disparity has been interpreted as
%signature of the competition of FM and AF
%fluctuations~\cite{IshidaK:YbRsft}.

Interestingly, $-(dM/dT)/T$ for $H\geq0.2$\,T shows a maximum (cf.
inset of Fig.~2(a) for data at 0.3~T) whose position coincides well
with the temperature where the specific heat coefficient passes a
maximum, $T_{\rm max}$. Similar behavior in $C(T)/T$ has also been
observed in other heavy-fermion compounds such as
CeCu$_{5.9}$Au$_{0.1}$ and
Ce$_{0.8}$Y$_{0.2}$Cu$_2$Si$_2$~\cite{lohneysen:prl-94,bredl:prl-84}.
The origin of such maxima may be understood qualitatively by the
Zeeman splitting of the Kondo resonance~\cite{Schotte} which results
in a more polarized state.
%excitations from heavy FL state to a local moment state where the
%Kondo singlet formation is destroyed by thermal fluctuations.
Our results as well as those for CeCu$_{5.9}$Au$_{0.1}$ and
Ce$_{0.8}$Y$_{0.2}$Cu$_2$Si$_2$~\cite{lohneysen:prl-94,bredl:prl-84}
show that the maximum temperature in $C(T)/T$ increases linearly
with increasing field. %This behavior is expected since the FL state
%is
%stabilized in higher fields. %The fact that $-(dM/dT)/T$ shows a
%maximum at the same temperature as the specific heat coefficient
%suggests that the former quantity also captures such excitations.
%
%indicates that, upon increasing the magnetic field, a more polarized
%NFL state is established.
For the magnetic Gr\"uneisen ratio $\Gamma_{\rm mag}(T)$, the maxima
in $C/T$ and $-(dM/dT)/T$ cancel each other, leading to a monotonous
crossover from NFL to FL behavior at fields larger than 0.2~T
(Fig.~2(c)).

Before discussing details of $\Gamma_{\rm mag}(T)$ data, we first
summarize the important theoretical conclusions most related to the
present study with respect to the degree of assumptions.
\\(1) $\Gamma_{\rm
mag}$ diverges in the approach of any field-induced QCP whenever the
characteristic energy scale of a system is continuously suppressed
to absolute zero~\cite{ZhuLijun:UnidGp}.
\\(2) Assuming scaling, the critical behavior is
governed by the correlation length $\xi$ and the correlation time
$\xi_{\tau}$ such that the temperature dependence in the quantum
critical regime $\Gamma_{\rm mag}(T,H=H_c)\propto T^{-1/\nu z}$
($\nu$ and $z$ are the correlation-length exponent, and dynamical
critical exponent, respectively). For the field dependence in the FL
regime, scaling analysis remarkably predicts not only a universal
functional dependence but even its {\it prefactor} without any
adjustable parameter, i.e. $\Gamma_{\rm
mag}(T=0,H)$=$-G_r(H-H_c)^{-1}$, with $G_r=\nu(d-z)$ (d:
dimensionality of the critical
fluctuations)~\cite{ZhuLijun:UnidGp,GarstM:SigctG}. Furthermore,
this prefactor equals the exponent in the divergence of the
Sommerfeld coefficient
$\gamma\propto(H-H_c)^{G_r}$~\cite{ZhuLijun:UnidGp,GarstM:SigctG}.
Thus, the character of the QCP is completely determined by the
values of $\nu$, $z$ and $d$.
\\(3) Within the itinerant theory, the correlation length exponent
$\nu=1/2$ and the dynamical critical exponent $z$ equals 2 and 3 for
AF and FM case, respectively~\cite{Millis}.

We now check the consistency of our experimental results with these
theoretical predictions. Obviously $\Gamma_{\rm mag}$ diverges as a
function of temperature at $H$=$H_c$=0.06\,T (Fig.~2(c)) proving the
existence of a field-induced QCP (assumption 1). The saturation
values of $\Gamma_{\rm mag}$ for $T\rightarrow 0$ as a function of
field, plotted in the inset of Fig.~2(c), diverge like
$(H-H_c)^{-1}$ as expected from scaling (assumption 2). The fit
reveals $-G_r(H-H_c^{fit})^{-1}$ with
$H_c^{fit}$=(0.065$\pm$0.004)\,T and $G_r=-0.30\pm0.01$. For
YbRh$_2$(Si$_{0.95}$Ge$_{0.05}$)$_2$, the field dependence of
$\gamma$ within the FL regime has been studied in detail revealing
$\gamma\propto(H-H_c)^{-0.33}$~\cite{CustersJ:brehea} which has an
exponent very close to our value of $G_r$. This underlines the
thermodynamic consistency of the data. We note also that the
obtained value of $G_r$ sets strong constraints for the scaling
parameters $\nu$, $d$ and $z$. Within the itinerant theory
(assumption 3), in which $\nu=1/2$, the observed $G_r$ would require
$d-z=2/3$ which is impossible. By contrast, a critical Fermi surface
model which may be relevant for a Kondo-breakdown QCP has been
proposed~\cite{Senthil}, in which the electronic criticality is
described by $\nu=2/3$, $z=3/2$, and $d=1$, yielding $G_r=-1/3$
similar is in our experiments.

\begin{figure}[t]
\includegraphics[width=0.8\linewidth,keepaspectratio]{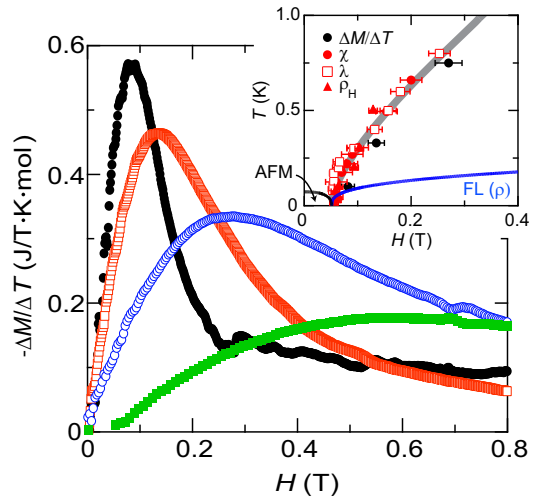}
\caption{(Color online) Magnetization difference divided by
temperature increment $-\Delta M/\Delta T=-\{M(T+\Delta
T,H)-M(T-\Delta T,H)\}/2\Delta T$ vs magnetic field for
YbRh$_2$Si$_2$, obtained using isothermal magnetization measurements
at $T=0.08$~K (black circles), 0.33~K (red open squares), 0.75~K
(blue open circles) and 1.5~K (green squares). Inset: $H$-$T$ phase
diagram of YbRh$_2$Si$_2$ with the peak positions in $-\Delta
M/\Delta T$ (black solid circles) and $T^{\star}(H)$ (grey solid
line) as determined from transport and thermodynamic properties (red
symbols)~\cite{PaschenS:Haleah,GegenwartP:Mulesa}. The black and
blue lines represent the phase boundary of the antiferromagnetic
(AFM) ground state and crossover to the Fermi liquid (FL) regime,
respectively~\cite{GegenwartP:Mulesa}.}
\end{figure}

Next, we focus on the temperature dependence of the magnetic
Gr\"uneisen ratio within the quantum critical regime. Interestingly,
divergent behavior is found not only at $H_c$ but also at 0.1~and
0.15~T (see Fig.~2c).
%From the first theoretical conclusion, $\Gamma_{\rm mag}(T)$
%diverges only at the critical field, when the characteristic
%temperature, in this case $T_{\rm N}$, goes to zero. It seems
%inconsistent with the theory that $\Gamma_{\rm mag}(T)$ diverges not
%only at $H$=0.06\,T but also at $H$=0.1, and 0.15\,T.
However, the obtained $H_c^{fit}$=0.065\,T is very close to 0.06~T,
supporting the QCP at $H_c$=0.06~T. Furthermore, the extrapolated
values of the saturated $\Gamma_{\rm mag}$ at $H$=0.1\,T and 0.15\,T
from $-G_r(H-H_c^{fit})^{-1}$ are 11.8 and 26.6\,T$^{-1}$,
respectively, strongly exceeding the largest {\it measured} values
shown in Fig.~2(c). Thus, the lowest temperature in our study
($\sim$70\,mK) is not low enough to observe saturation of
$\Gamma_{\rm mag}$ at fields very close to the QCP and thereby
explains the absence of saturation in $-(dM/dT)/T$ at 0.1 and
0.15~T.
%Therefore, we conclude that our results are consistent with the
%above mentioned theoretical predictions.
As shown in the double-log plot in the inset of Fig.~2(b),
$\Gamma_{mag}\propto T^{-\epsilon}$ with $\epsilon\approx 2.0$ from
3~K down to 0.3~K. This temperature dependence is much stronger than
any expectation within the itinerant theory. At lower temperatures,
a crossover to a weaker divergence is found with $\epsilon\approx
0.7$ below 0.25~K, which interestingly is similar as the exponent
obtained by the study of the thermal Gr\"{u}neisen ratio for
YbRh$_2$(Si$_{0.95}$Ge$_{0.05}$)$_2$ at zero
field~\cite{KuchlerR:DivtGr}. The crossover around 0.3~K may be
related to the interplay of AF and FM fluctuations, as it is of the
same order of magnitude as the absolute value of the Weiss
temperature~\cite{GegenwartP:FM}. The fact that no single power-law
behavior is found over a larger temperature regime makes the
interpretation of the Gr\"uneisen ratio within the quantum critical
regime difficult.
%Since the absolute value of the obtained Weiss temperature ~-0.46\,K
%is comparable to the crossover temperature, it is reasonable to
%assume that the lower-temperature exponent $\epsilon\approx 0.7$ is
%responsible to the critical fluctuations.
%The critical Fermi surface model for a local QCP has recently been
%proposed~\cite{Senthil}, which partly explains our results. In the
%theory, the electronic criticality is described by $\nu=2/3$,
%$z=3/2$, and $d=1$, yielding $G_r=-1/3$, in agreement with our
%experiments. However it gives $\epsilon$=1, slightly larger than the
%observed exponent. Up to now there is no theoretical model which
%provides both the obtained values of $G_r$=-0.3 and the temperature
%exponent $\epsilon$=0.7. A rather complicated situation in this
%compound with the competition between AFM and FM fluctuations may be
%responsible to this deviation.

At last, we study the evolution of the magnetic entropy upon
crossing $T^\star(H)$. For this purpose, it is important to follow
the field dependence of $dM/dT$, which according to the Maxwell
relation equals the field dependence of $dS/dH$. Using isothermal
magnetization scans at different temperatures, we calculate the
difference in magnetization divided by the temperature increment,
$-\Delta M/\Delta T=-\{M(T+\Delta T,H)-M(T-\Delta T,H)\}/2\Delta T$
vs magnetic field, plotted in Fig.~3. For each curve, we have used
two sets of isothermal magnetization $M(H)$ data at temperatures
$T-\Delta T$ and $T+\Delta T$. The $-\Delta M/\Delta T(H)$ traces
show a peak whose position shifts to larger field with increasing
temperature as displayed in the temperature versus field phase
diagram in the inset of Fig.~3.
%Extrapolation by third polynomial fitting gives $H_{\rm
%peak}(T=0)$=0.06\,T, consistent with the divergence of $dM/dT$ at
%the QCP. Furthermore, t
The peak positions agree satisfactorily with the energy scale
$T^\star(H)$ found previously in the Hall coefficient
\cite{PaschenS:Haleah}, magnetostriction and susceptibility
\cite{GegenwartP:Mulesa} (cf. red symbols in the inset of Fig.~3).
%In the inset of Fig.~4, the position of the anomalies due to the
%additional field scale $H^{\star}$ (equivalent to the temperature
%scale $T^{\star}$) found in susceptibility $\chi$, magnetostriction
%$\lambda$ and Hall resistivity $\rho_{\rm H}$ are plotted for
%comparison~\cite{PaschenS:Haleah,GegenwartP:Mulesa}. Interestingly,
%the peak position of $-\Delta M/\Delta T$ agrees satisfactorily with
%the field scale $H^{\star}$.
Through $\Delta M/\Delta T\approx dM/dT=dS/dH$, the field dependence
of the entropy $S(H)$ could be obtained by integrating $\Delta
M/\Delta T$ over the field. Since $dM/dT<0$ in the entire field
range (at all accessible temperatures), the magnetic entropy
decreases with increasing field. The positions of the maxima in
$-\Delta M/\Delta T$ correspond to inflection points in $S(H)$, i.e.
they mark characteristic magnetic fields, at which entropy $S(H)$ at
constant temperature is reduced most strongly with increasing field.
Remarkably, these inflection points of the entropy agree very well
with $T^\star(H)$ determined from Hall
effect~\cite{PaschenS:Haleah}, magnetostriction and
magnetization~\cite{GegenwartP:Mulesa} and thus provide further
thermodynamic confirmation of this additional energy scale in
YbRh$_2$Si$_2$.

%The integrated $\Delta M/\Delta T(H)$ agrees well with the $S(H)$
%values obtained by integrating $C(T)/T$ data reported in Ref.[17].
%Since the position of this signature in the phase diagram agrees
%well with $T^{\star}(H)$, we consider it a as further thermodynamic
%confirmation of this novel scale.

%The power law exponent $\epsilon$ for the magnetic Gr\"{u}neisen
%ratio is mapped in the $H$-$T$ phase space as shown in the contour
%plot (Fig.~4). The colors represent the exponent. At temperatures
%above ~0.25\,K, the map of exponents behaves as expected. The
%contour forms an upside-down cone shape centered at the critical
%field, and the exponent becomes zero below $T_{\rm max}$. However,
%as discussed earlier, below ~0.25\,K, there is a region where
%$\Gamma_{\rm mag}(T)$ ($-(dM/dT)/T$) diverges, while $C/T(T)$ is
%constant (area with black solid lines in Fig.~4). This may be due to
%different cross-over temperatures in $C/T$ and $-(dM/dT)/T$.

In conclusion, our study of the low-temperature magnetization and
specific heat of YbRh$_2$Si$_2$ reveals for the first time in any
system a divergence of the magnetic Gr\"uneisen ratio $\Gamma_{\rm
mag}=-(dM/dT)/C$ at a magnetic field-tuned QCP. This property
provides information on the scaling of the magnetic entropy due to
quantum critical fluctuations. The field-dependence of the magnetic
Gr\"uneisen ratio within the FL regime follows $\Gamma_{\rm
mag}(T\rightarrow 0)=-G_r(H-H_c)^{-1}$~\cite{ZhuLijun:UnidGp} with a
prefactor $G_r=-0.3\pm 0.01$ that is consistent with a recent
critical Fermi surface model~\cite{Senthil}. Within the quantum
critical regime, complicated behavior with a crossover scale around
0.3~K is found which may be related to the interplay of AF and FM
fluctuations in this system.
%To conclude, we have measured the DC magnetization and specific heat
%of YbRh$_2$Si$_2$ for $H\perp c$ at and above the quantum critical
%field $H_c$=0.06\,T and obtained the magnetic Gr\"{u}neisen ratio
%$\Gamma_{\rm mag}=-(dM/dT)/C$. Our results prove the first-ever
%divergence of the magnetic Gr\"{u}neisen ratio and are used to
%
%We have observed, for the first time, a divergence of $\Gamma_{\rm
%mag}(T)$ at the field-induced QCP in YbRh$_2$Si$_2$. The field
%dependence of saturated $\Gamma_{\rm mag}$ in the FL regime follows
%the theoretical prediction $\Gamma_{\rm mag}(T\rightarrow
%0)=-G_r(H-H_c)^{-1}$~\cite{ZhuLijun:UnidGp} with the prefactor
%$G_r=-0.3\pm 0.01$ which is thermodynamically consistent with the
%previous specific-heat results on the slightly Ge-doped system.
%We also found separated behaviors between $C/T$ and $-(dM/dT)/T$ for
%0.1\,T$\leq H\leq$0.15\,T; i.e., saturated $C/T$ and diverging
%$-(dM/dT)/T$. This results in diverging $\Gamma_{\rm mag}$ not only
%at 0.06\,T, but also at 0.1 and 0.15\,T, however, $(H-H_c)^{-1}$
%fitting for saturated $\Gamma_{\rm mag}(T\rightarrow 0)$ indicates
%that the saturation temperature at those fields is lower than the
%lowest temperature we have studied (~70\,mK). Thus, our result does
%not contradict with the main theoretical prediction, i.e., a
%diverging $\Gamma_{\rm mag}$ only at $H$=$H_c$.
We also observe an inflection point in the entropy at the scale
$T^\star(H)$ previously observed in Hall- and thermodynamic
measurements~\cite{PaschenS:Haleah,GegenwartP:Mulesa}. A reduction
of spin entropy when entering the entangled heavy-electron fluid
upon increasing the magnetic field is consistent with the
delocalization of $f$-electrons upon field tuning through the QCP.

We would like to acknowledge helpful discussions with M. Brando, M.
Garst, R. Movshovich, N. Oeschler, F. Ronning, Q. Si, and T.
Senthil. This work was supported by the DFG research unit 960
"Quantum phase transitions".

%\begin{thebibliography}{10}

%\end{thebibliography}
\end{document}